\documentstyle[11pt,a4]{article}

\newcommand {\beq}{\begin{equation}}
\newcommand {\eeq}{\end{equation}}

\newcommand {\spin}{\mbox{\small$\frac{\pi}{2}$}}
\newcommand {\Tr}{\mathop{{\rm Tr}}}

\newcommand{\be}{\begin{equation}}
\newcommand{\ee}{\end{equation}}
\newcommand{\bea}{\begin{eqnarray}}
\newcommand{\eea}{\end{eqnarray}}
\newcommand{\bean}{\begin{eqnarray*}}
\newcommand{\eean}{\end{eqnarray*}}
\newcommand{\stackrange}[2]{{\scriptstyle #1 \atop \scriptstyle #2}}

\newcommand{\vac}{|0\rangle}

\newcommand{\gapproxeq}{\lower .7ex\hbox{$\;\stackrel{\textstyle >}{\sim}\;$}}
\newcommand{\lapproxeq}{\lower .7ex\hbox{$\;\stackrel{\textstyle <}{\sim}\;$}}
\newcounter{appendice}
\newcommand{\appendice}
{
\setcounter{equation}{0}
\renewcommand{\theequation}{\Alph{appendice}.\arabic{equation}}
\addtocounter{appendice}{1}
{\bf Appendix \Alph{appendice}}
}

\def\thebibliography#1{\section*{REFERENCES\markboth
 {REFERENCES}{REFERENCES}}\list
 {[\arabic{enumi}]}{\settowidth\labelwidth{[#1]}\leftmargin\labelwidth
 \advance\leftmargin\labelsep
 \usecounter{enumi}}
 \def\newblock{\hskip .11em plus .33em minus -.07em}
 \sloppy
 \sfcode`\.=1000\relax}

\def\thebibliography#1{\section*{REFERENCES\markboth
 {REFERENCES}{REFERENCES}}\list
 {[\arabic{enumi}]}{\settowidth\labelwidth{[#1]}\leftmargin\labelwidth
 \advance\leftmargin\labelsep
 \usecounter{enumi}}
 \def\newblock{\hskip .11em plus .33em minus -.07em}
 \sloppy
 \sfcode`\.=1000\relax}

\renewcommand{\theequation}{\thesection.\arabic{equation}}

\renewcommand{\theequation}{\thesection.\arabic{equation}}

\begin{document}
\begin{titlepage}
\begin{flushright}
FTUV \\
IFIC \\
\end{flushright}
\vskip 1cm
\begin{center}

{ \Large \bf
N-STRING VERTICES IN STRING FIELD THEORY} \\
\vskip 1cm
{\bf \large J.  BORDES\footnote{Also at IFIC, Centro Mixto Universitat de
Valencia-CSIC. Spain.} } \\
 \vskip 0.2cm
 {\it Departament de
F\`{\i}sica Te\'{o}rica. Universitat de Val\`{e}ncia. \\
C./ Dr. Moliner 50.
E-46100. Burjassot (Val\`encia) Spain.}, \\
\vskip 0.6cm
{\large A. ABDURRAHMAN\footnote{Also at Mathematical Institute, 24-29, St.
Giles, Oxford, OX1 3LB, U.K.} and  F. ANT\'ON} \\
\vskip 0.2cm
{\it Department of Theoretical Physics, University of Oxford} \\
{\it 1 Keble Road, Oxford, OX1 3NP, U.K.}

\end{center}

\begin{abstract}
We give the general form of the vertex corresponding to the interaction
of an arbitrary number of strings. The technique employed relies on the
``comma" representation of String Field Theory where string fields
and interactions are represented as matrices and operations between them
such as multiplication and trace. The general formulation presented here
shows that the interaction vertex of
N strings, for any arbitrary N, is given as a function
of particular combinations of matrices corresponding to the change of
representation between the full string and the
half string degrees of freedom.
\end{abstract}
\begin{flushleft}
January 1993
\end{flushleft}
\end{titlepage}

\section{INTRODUCTION}

String Field Theory has provided a consistent
picture for the treatment of Open Strings \cite{WITTEN}.
Recently some advances have been made in the formulation of
a Field Theory for Closed Strings \cite{ZWIEBACH}. This is
welcomed from
a phenomenological point of view since closed strings
appear to give a
suitable picture of string physics at low energy. A complete understanding
of low energy string physics seems to require the treatment
of strings in this framework.

The Closed String Field Theory (CSFT) proposed in \cite{ZWIEBACH} has the
particularity of requiring a non-polynomial action in which at every
step one has to include a term in the action corresponding to the interaction
of an arbitrary number of strings over a world sheet given by the so-called
restricted polyhedra. The edges of these polyhedra play the role of the modular
parameters and one restricts the region of integration over these in a
prescribed way. In the theory, the interaction terms are interpreted as the
overlapping of closed strings in a way which resembles
the original theory for open strings due to Witten \cite{WITTEN}.
On the other hand there is the
suggestion in \cite{CHANII} that the overlapping of closed strings can be
formulated, using the property of reparametrization invariance of the string
amplitudes, as the overlapping of standard string segments which, following the
example of the open strings, can be considered as half-strings. In turn,
one interprets the string functionals as matrices and the interaction between
strings as the product and trace of these matrices.
This feature shows that, even in the case of closed strings, the half-string
picture is relevant in the construction of the string interaction. However, the
formulation from a Fock space approach seems to be a formidable task since in
the absence of a compact formulation one should calculate separately every term
in the action.

Hence, our main motivation comes from the fact that,
following \cite{CHANII},
the $N$-faced polyhedra describing the interaction of $N$ strings
at the level of the action, can be written as a reparametrization
of the vertex in a contact interaction.
Thus, as a first step, one finds useful to work out the contact
interaction vertex for an arbitrary number of strings.

The purpose of this work is then the study of the $N$--strings
contact interaction
in a general form. This will form the base of a compact formulation
of the closed
string vertex in the non-polynomial theory. We perform the calculation
for open strings since their formulation at the level of Fock
space is firmly settled. On the other hand, they share many
of the peculiarities
of the operator formulation for closed strings \cite{CLOSED}.
The results obtained here are very similar --from the
technical point of view-- to those for the closed string so they can
be adapted, with minor modifications,
to the latter.

Related to this point, we will see that, apart from the fact that half-strings
play a conceptual role in the formulation of closed string interactions,
linking the CSFT \cite{ZWIEBACH} with the approach described in
\cite{CHANII}, they reveal as
a useful technical tool for the treatment of string amplitudes, both for
open and closed strings.

In the theory of Witten \cite{WITTEN} for string fields the
interaction between strings is defined by a
$\star$ (product) and  $\int$ (integration) which are
defined through the joining of half strings. In particular, the
interaction between two strings to give a third one is defined
from the $\star$ as:
\be
(\psi \ast \phi) [{\bf x}] = \int D{\bf y} \psi[{\bf x}_L ; {\bf y}
] \phi [ \overline{{\bf y}} ; {\bf x}_R],
\label{STAR}
\ee
where $\psi$ represents the string fields.
The interaction takes place through the joining of half
strings.The operation of integration, which allows us to obtain invariant
quantities, is defined as:
\be
\int \psi = \int D{\bf x} \psi[{\bf x} ;\overline {{\bf x}}].
\label{WINT}
\ee
Therefore the fundamental degree of freedom from
the interaction point of view is the half string.
It is then justified to adopt an approach in which
the half string plays explicitly an important role.
We will base our formalism in the ``comma" representation of string
field theory developed in \cite{JOSE}.

In the ``comma" formulation, we single out the
midpoint and represent string fields as matrices. The string is
divided into left and right parts which play the role of row and
column indices of those matrices. Interaction takes place
simply by multiplying (product (\ref{STAR}))
and taking the trace (integration (\ref{WINT})) over
the matrices. The advantage
of this approach is that one can handle
in a compact form the $N$ string contact interaction for
any arbitrary $N$. In fact, the vertex for $N$ strings is simply given
by:
\be
V_N = \int d {\bf x}(\spin) \; \; \Tr[A_1 . A_2 \ldots A_N],
\label{INTERACTION}
\ee
where $A_i$ are the matrices representing the string states.

We must point out that the former equation does not give the
tree level $N$-string interaction, but only the contact term.
In order to get the tree level amplitude one could consider
the reparametrization approach as described in \cite{CHANII}
to get the correct moduli space of parameters. This
however, is beyond the scope of this paper, although work in
this direction is under way in order to find a formulation
suitable for the non-polynomial CSFT.

To finish the introduction some comments are necessary about
the ghost degrees
of freedom. It is known that in Witten's theory, the violation of
ghost number at the vertices and the ghost number of
physical states fixes the value of $N$.
In this sense, the vertex (\ref{INTERACTION}) vanishes unless $N=3$.
This is of no relevance to us since our purpose is just the
calculation of the $N$ string interaction vertex in order
to get some insight of the structure of the terms in the theories
of \cite{ZWIEBACH}. Hence in the following we will ignore the ghosts
degrees of freedom, although they certainly could be treated
with the techniques presented here~\cite{GHOSTS}.

The plan of the paper is as follows. In section 1 we present the
formulation of the half string degrees of freedom (``comma"
representation). In section 2 we discuss the construction of the
string physical states, therefore settling the
basis for the calculation of the
$N$ strings vertex. This calculation is performed in section 3.
In section 4 we find the Fourier coefficients of the Neumann
functions for the appropriated geometry of the vertex. Comparison
with particular cases worked out previously is also given. Finally
we summarize our results in the conclusions.

\section{STRING AND ``COMMA" COORDINATES}

The first task we must face is the expression of
the elements of string field theory, namely the fields and
the interactions between them, in terms of matrices and operations such
as trace and multiplication (\ref{INTERACTION}). To this end
one has to introduce the
degrees of freedom referring to the left and right halves of the string.
Following our previous work \cite{JOSE} we define the functions:
\bea
\chi^{(1)}(\sigma,\tau) & = & {\bf x}(\sigma , \tau)  - {\bf x}
(\spin,\tau ),
\nonumber \\
\chi^{(2)}(\sigma,\tau) & = & {\bf x}(\pi-\sigma , \tau)  - {\bf x}
(\spin,\tau ),
\nonumber \\
\quad\quad\quad \sigma & \in & [0,\spin],
\label{COORDINATES}
\eea
where ${\bf x}(\sigma , \tau)$ are the string coordinates
(space-time indices will be suppressed throughout). From their
definition, it is clear that $\chi^{(1)}$ and $\chi^{(2)}$ are
related to the left and right parts of the string respectively,
(henceforth we will call them ``comma"
coordinates).

The boundary conditions as well as the constraint implied in the
change of representation (see \cite{JOSE} for details) imply the
Fourier expansion of the ``comma" functions in terms of odd cosine modes with
the explicit form:
\begin{equation}
\chi^{(r)}(\sigma,\tau) = \sqrt{2} \sum_{n \geq 1} \; \chi^{(r)}_n(\tau) \;
\cos(2n-1) \sigma
\quad {\rm where}
\quad r=1,2; \; \; \sigma \in [0,\spin].
\nonumber
\end{equation}
The inverse relations are obtained by integration over one period:
\be
\chi^{(r)}_n (\tau) = \frac{2\sqrt 2}{\pi} \int^{\pi/2}_0 d\sigma \; \chi^{(r)}
(\sigma, \tau) \; \cos (2n-1)\sigma
\quad{\rm where}
\quad r=1,2.
\label{FOUINV}
\ee

Our purpose now is to find the relation between the ``comma" modes and
the conventional string ones. This will allow us to represent
the physical states in terms of the ``comma" degrees of freedom.

{}From definitions  (\ref{COORDINATES}) and (\ref{FOUINV}),
using the standard open string mode expansion:
\be
{\bf x}(\sigma,\tau) = x_0 + p\tau + i \sum_{n\neq 0} \frac{\alpha_n}{n}
e^{-in\tau} \cos n\sigma,
\nonumber
\ee
one arrives at the relation:
$$
\chi^{(r)}_n(\tau)  = \sum_{m\neq 0} \chi^{(r)}_{n,m} e^{-im\tau},
$$
where the $\chi^{(r)}_{n,m}$ are time-independent coefficients given
in terms of string oscillator modes by:
\bea
\chi^{(r)}_{n,2m} & = &  -\sqrt{2} \frac{(-)^{n+m}}{\pi}
\left(
\frac{1}{2m+2n-1} -
\frac{1}{2m-(2n-1)} -\frac{2}{2n-1}
\right)
\frac{i\alpha_{2m}}{2m},
\nonumber \\
\chi^{(r)}_{n,2m-1} & = & 0,
\nonumber
\eea
for $n\neq m,-m$ and
\bea
\chi^{(r)}_{n,n} & = & \frac{(-)^{r+1}}{\sqrt{2}} \frac{i\alpha_{2n-1}}{2n-1},
\nonumber \\
\chi^{(r)}_{n,-n} & = &
   \frac{(-)^{r}}{\sqrt{2}} \frac{i\alpha_{-(2n-1)}}{2n-1}.
\label{TRANSFORM}
\eea

Since we will mainly deal with string fields, namely string wave functionals,
we are only interested in the relations at fixed $\tau$.  Fixing $\tau = 0$ in
(\ref{TRANSFORM}) we end up with a relation between the ``comma" and the string
oscillator modes, the latter defined according to:
\bea
x_m = \frac{i}{\sqrt{2} m} (\alpha_m - \alpha_{-m}),
\nonumber
\eea
to give:
\be
\chi^{(r)}_n = (-)^{r+1} x_{2n-1} + \sum_{m \geq 1}
\left(
\frac{2m}{2n-1}
\right)^{1/2}
\left[
(M_1)_{m,n} + (M_2)_{m,n}
\right]
x_{2m}; \; \; \quad  r= 1,2,
\label{TRANSFORMATION}
\ee
where the matrices $M_1$ and $M_2$ are:
\bea
(M_1)_{n,m} & = & \frac{2}{\pi}
\left(
\frac{2n}{2m-1}
\right)
^{1/2}
\frac{(-)^{n+m}}
{2n-(2m-1)}  ,
\nonumber \\
(M_2)_{n,m} & = & \frac{2}{\pi}
\left(\frac{2n}{2m-1}
\right)
^{1/2}
\frac{(-)^{n+m}}
{2n+2m-1} .
\label{MATRICES}
\eea

It has been shown that the transformation (\ref{TRANSFORMATION}) is
non-singular \cite{JOSE}, the
inverse relation can be obtained from (\ref{FOUINV}) to give:
\bea
x_{2n-1} & = &
\frac{1}{2} \left(
\chi^{(1)}_n - \chi^{(2)}_n
\right)
\nonumber \\
x_{2n} & = &
\frac{1}{2}
\sum_{m\geq1}
\left(
\frac{2m-1}{2n}
\right)^{1/2}
\left[
(M_1)_{m,n} - (M_2)_{m,n}
\right]
\left(
\chi^{(1)}_m + \chi^{(2)}_m
\right).
\label{INVERSE}
\eea

In the decomposition of the string into left and right halves
(\ref{COORDINATES}), we have singled out the midpoint coordinate.
Therefore to
complete the picture we need its expression in string coordinates
which, at $\tau =0$, reads:
\bea
{\bf x}(\spin) = x_0 +\sqrt{2} \sum_{n\geq1} (-)^n x_{2n},
\nonumber
\eea
conversely, the center of mass in the ``comma"
representation is (again we take $\tau = 0$):
\bea
x_0 = {\bf x}(\spin) - \frac{\sqrt{2}}{\pi}
\sum_{\stackrange{r=1,2}{n\geq1}} \frac{(-)^n}{2n-1} \chi^{(r)}_n .
\nonumber
\eea

These two relations together with the oscillator mode relations (\ref
{TRANSFORMATION}, \ref{INVERSE}) complete the equivalence between the string
oscillator modes $\{x_n\}^\infty_{n=0}$ and the position degrees of freedom
describing the ``comma" representation, which, in the transformation we have
defined, are equivalent to the midpoint and the ``comma" oscillator
modes, i.e., $\{\chi^{(r)}_n;{\bf x}
(\spin)\}$ where $r=1,2$ and $n=1,\ldots, \infty$.

\subsection{Conjugate Momentum}

For our purpose, we merely need the relations between the ``comma" and
string conjugate momenta, we can define
the quantized momentum conjugate to $\chi^{(r)}_n$ and {\bf x}(\spin)
in the usual way:
\bea
{\cal P}^{(r)}_n = -i \frac{\partial}{\partial\chi^{(r)}_n},
\; \; \; {\cal P} = -i \frac{\partial}{\partial{\bf x}(\spin)}.
\nonumber
\eea
Thus, using (\ref{TRANSFORMATION}) and applying the chain rule, one can find
the
relation with the conventional string momenta $(p_m)$.
In summary, they are given by:
\bea
{\cal P}^{(r)}_n & = & \frac {1}{2} p_{2n-1} +
\frac {1}{2} \sum_{m\geq1}
 \left( \frac{2n-1}{2m} \right)^{1/2}
\left[
(M_1)_{m,n} + (M_2)_{m,n}
\right] p_{2m}
- \frac{\sqrt{2}}{\pi} \frac{(-)^n}{2n-1} p_0 ,
\nonumber \\
{\cal P} & = & p_0.
\nonumber
\label{MOMENTA}
\eea
Also, the inverse relations
read:
\bea
p_{2n-1} & = & {\cal P}^{(1)}_n - {\cal P}^{(2)}_n \\
p_{2n}   & = &
\sum_{m\geq1} \!
\left( \frac{2n}{2m-1} \right)^{1/2}
\left[
(M_1)_{n,m} + (M_2)_{n,m}
\right]
\left(
{\cal P}^{(1)}_m + {\cal P}^{(2)}_m
\right) +
\nonumber \\
& + &
\sqrt{2} (-)^n {\cal P}.
\nonumber
\label{MOMENTABIS}
\eea

Upon quantization, the commutation relations for the ``comma" coordinates
and momenta
are the usual ones corresponding to a discrete set
of conjugate variables, namely:
\bea
[\chi^{(r)}_n,{\cal P}^{(s)}_m] & = & i \; \delta^{rs} \delta_{mn},
\nonumber \\
\left[{\bf x}(\spin),{\cal P}\right] & =&  i,
\nonumber
\eea
as can be explicitly checked from the previous relations.

{}From the one to one correspondence between
the ``comma" and the string degrees of freedom as we have established in
(\ref{TRANSFORMATION})-(\ref{MOMENTABIS}),
we see that the identification
$$
\overline{{\cal H}}=
\overline{{\cal H}_M \otimes {\cal H}_1 \otimes{\cal H}_2},
$$
is needed
where ${\cal H}$ stands for the string space, ${\cal H}_r$
are two copies of the half-string spaces and ${\cal H}_M$ describes the
midpoint. The over bar in the former expression stands for the
completion of spaces, we need to take this completion
in order to ensure a Hilbert space structure in both
terms.

\section{OPERATOR APPROACH TO THE COMMA REPRESENTATION}

For practical purposes it is convenient to develop a formulation based
on creation and annihilation operators and express the elements of the
theory in terms of the tensor product of two copies of the Fock space
representing the states of each half-string.
This will allow us to
overcome the ambiguities in the definition of the functional integral
appearing in the product (\ref{STAR}) and integration (\ref{WINT}), since
one ends up with a representation in which the states are just infinite
matrices
and the above-mentioned operations become
products and traces of matrices respectively.

In order to construct the Fock space of the ``comma" states, let us define the
creation and annihilation operators for the ``comma" modes in the usual way:
\bea
b^{(r)}_n = \frac{-i}{\sqrt{2}}
\left(
\frac{2n-1}{2}
\right)
^{1/2}
\left\{
\chi^{(r)}_n + i\frac{2}{2n-1} {\cal P}^{(r)}_n
\right\} ,
\nonumber \\
b^{(r)\dagger}_n = \frac{i}{\sqrt{2}}
\left(
\frac{2n-1}{2}
\right)
^{1/2}
\left\{
\chi^{(r)}_n - i\frac{2}{2n-1} {\cal P}^{(r)}_n
\right\} ,
\nonumber \\
(n \geq 1).
\nonumber
\eea

The degrees of freedom relative to the
midpoint, namely ${\bf x}(\spin)$ and $\cal P$;
only appear in the Fock space states in the form of a plane wave.
We can use directly those variables
to generate the midpoint Hilbert space.

The meaning of the operators $b^{(r)\dagger}_n$ is clear:
acting on the ``comma" vacuum state
$\vac_r$, creates a ``comma" oscillator mode of half integral
 frequency $(n-\frac{1}{2})$,
these vacua fulfill the relation $b^{(r)}_n \vac_r = 0$, $n=1,\ldots,
\infty$.

Repeated action of $b^{(r)\dagger}_n$ on the vacuum gives the Fock space
states corresponding to each half-string $({\cal H}_r)$.  Hence, the complete
space on which the ``comma" states live is given by the tensor product
$\hbox{\boldmath $\cal H$} =
{\cal H}_1 \otimes {\cal H}_2 \otimes {\cal H}_M$ after folding in the
piece corresponding to the midpoint motion (${\cal H}_M$).

Following similar steps as in the preceding section, we can find
the relation
between the ``comma" creation and
annihilation operators and the conventional ones.  Using equations
(\ref{TRANSFORMATION}, \ref{MOMENTA}) we find:
\bea
b^{(r)}_n =
\frac{\sqrt{2}}{\pi}
\frac{(-)^{(n-1)}}{(2n-1)^{3/2}} p_0
+ \frac{1}{\sqrt{2}}
(-)^{r+1} a_{2n-1}+
\sum^\infty_{m=1}
\left[
(M_1)_{m,n} a_{2m} -
(M_2)_{m,n} a_{2m}^\dagger
\right],
\nonumber \\
\label{BETA}
\eea
and the corresponding relation for $b^{(r)\dagger}_n$ by changing $a_n
\rightleftharpoons a_n^\dagger$.

Again, the inverse relations are given by:
\bea
a_{2n-1} &  =  & b_n^{(-)},
\nonumber \\
a_{2n} & = &
\frac{(-)^n}{\sqrt{2n}} {\cal P}
+ \sum_{m\geq 1}
\left[
(M_1)_{n,m} b_m^{(+)}
- (M_2)_{n,m} b_m^{(+) \dagger}
\right].
\nonumber \\
\label{ALFA} \\
\nonumber
\eea
We have defined the combinations $b_m^{(\pm)} = \frac{1}{\sqrt 2}
\left( b_m^{(1)} \pm b_m^{(2)} \right) $ --the ones corresponding to
the
creation operators $b_m^{(r)\dagger}$ are defined in an analogous fashion--.
Finally the relation for $a_n^\dagger$ is accordingly
obtained.

With relations (\ref{BETA}, \ref{ALFA}) in hand, we are able to obtain the
string states in terms of the
tensor product of ``comma" Fock space
states; in particular we find that they belong to the completion
of the space
{\boldmath ${\bf \cal H}$} defined above.

\subsection{String States in the ``Comma" Representation}

It was already mentioned that string states,
as gauge invariant states and eigenstates of the string
Hamiltonian \cite{GREENBOOK},
are defined from the conventional string creation and
annihilation operators.
In the following we use this definition and the relation with the
``comma" modes (\ref{ALFA}) to write them in the ``comma"
representation.

First of all we shall start with the string vacuum which,
in the string representation, is defined
through the
relations $a_n \vac =0  \; \forall n$.  Then in view of relations
(\ref{ALFA}) we can
express the vacuum state as an exponential of a
quadratic form in creation operators
$b^{(r)\dagger}_n$ acting on the tensor product $\vac_1 \vac_2$.
Notice that $a_{n=odd} \vac = b^{(-)}_n  \vac =0$, so only
the combination
$b^{(+)^\dagger}_n$ appears in the exponential. Under this conditions the
vacuum takes on a generic form similar to the BCS vacuum involving only the
$b^{{(+)}\dagger}_n$ operators, namely:
\bea
\|0\rangle = \exp
\left(
-\frac{1}{2} b^{{(+)}\dagger}_n  \phi_{n,m} b^{{(+)}\dagger}_m
\right)
\vac_1\vac_2,
\label{VACUUMBIS}
\eea
and the matrix $\phi$ is determined by acting with $a_{n=even}$. One finds:
\bea
\phi = M_1^{-1} M_2,
\nonumber
\eea

Using the properties of the coefficients
given in the appendix, and noticing that in the particular case $p=2$
they are related to the combinatorial numbers $\left( \begin{array}{c}
-1/2 \\ n \end{array} \right)$,
one arrives at the following expression for the elements of
the matrix $\phi$:
\bea
\phi_{n,m} = (2n-1)^{1/2} (2m-1)^{1/2}
\frac{1}{2(n+m-1)}
\left(
\begin{array}{c}
-1/2 \\
n-1
\end{array}
\right)
\left(
\begin{array}{c}
-1/2 \\
m-1
\end{array}
\right).
\label{PHI}
\eea

The tachyon state is immediately obtained just by inserting the
plane wave of momentum $p$ corresponding to the center of mass motion.  Since
the center of mass operator is expressible in terms of creation and
annihilation operators as:
\bea
x_0 = {\bf x}(\spin) - i\frac{\sqrt{2}}{\pi} \sum_{n\geq1}
\frac{(-)^n}{(2n-1)^{3/2}} (b^{(+)}_n - b^{(+)\dagger}_n),
\nonumber
\eea
one gets the tachyon in the ``comma" representation:
\bea
\|T\rangle = e^{ip{\bf x}(\spin)}
\exp
\left[
-\frac{1}{2}p^2
 \vec{k}^T(I+\phi)\vec{k}-p\vec{k}^T(I+\phi)\vec{b}^{{(+)}\dagger}
\right]
\nonumber \\
\exp
\left(-\frac{1}{2}\vec{b}^{{(+)}\dagger}\phi \vec{b}^{{(+)}\dagger}
\right)
\vac_1 \vac_2 .
\label{TACHYON1}
\eea
The vector $\vec{k}$ is given by $k_n = \frac{2}{\pi} \frac{(-)^n}{(2n-1)
^{3/2}} $ and the operators $b^{(-)\dagger}_n,b^{(+)\dagger}_n$
are given above as a
combination of the ``comma" annihilation operators. As we can see,
the momentum
insertion unfolds into two pieces, the first one giving the midpoint motion,
the second relating the
two halves of the string.

\subsection{String States}

In the usual string picture,
higher states are obtained by the action on (\ref{TACHYON1}) of the creation
operators $a^\dagger_n$ (since $[a^\dagger_n, x_0]=0$ there is no ambiguity in
this construction).

To get the general setting it
is useful to work out the coherent state bases of the
string states defined by:
\bea
\|\vec{\lambda},\vec{\lambda}^\prime; p) & = &
e^{ipx_0} \|\vec{\lambda},\vec{\lambda}^\prime ) =
\nonumber \\
& =  &
e^{ipx_0}
\exp
\left(
\sum_{n=1}^\infty
(\lambda^\prime_n a^\dagger_{2n} + \lambda_n a^\dagger_{2n-1})
\right)
\|0 \rangle,
\label{COHERENT1}
\eea
in the ``comma'' representation.
By taking the appropriate derivatives of these states, one can obtain string
states with any occupation numbers.
In general, for the string state with occupation
numbers $\{n_i\}^\infty_{i=1}$ one has:
\bea
\| \{n_i\}^\infty_{i=1} \rangle = \prod^\infty_{i=1} \frac{1}{\sqrt{n_i!}}
(a^\dagger_i)^{n_i} \|0\rangle =
\nonumber \\
=\prod^\infty_{i=1} \frac{1}{\sqrt{n_i!}} \frac{\partial^{n_{2i-1}}}
{\partial\lambda_i^{n_{2i-1}}}
\frac{\partial^{n_{2i}}}
{\partial\lambda_i^{\prime n_{2i}}} \|\vec{\lambda},\vec{\lambda}^\prime).
\label{COHERENTFULL}
\eea

These states are written in the ``comma" representation making the
appropriate substitution of the string oscillator creation operators
in terms of the ``comma" ones (\ref{ALFA}). Moving the annihilation
operators to the right, one is left with the following expression
for the coherent string state:
\bea
\|\vec{\lambda},\vec{\lambda}^\prime; p ) =
{\bf C}(p,\vec{\lambda}^\prime)
e^{ip{\bf x}(\spin)}
\exp
\left(
\vec{\lambda}^T\vec{b}^{(-)\dagger} +\vec{\rho}^T\vec{b}^{(+)\dagger}
-\frac{1}{2}\vec{b}^{(+)\dagger}\phi\vec{b}^{(+)\dagger}
\right)
\vac_1\vac_2,
\nonumber \\
\label{COHERENT2}
\eea
where:
\bea
{\bf C}(p,\vec{\lambda}^\prime) & = &
\exp
\left(
-\frac{1}{2} p^2 \vec{k}^T(I+\phi)\vec{k} +
\frac{1}{2} \vec{\lambda}^{\prime T} (M_1^T)^{-1} M^T_2\vec{\lambda}^\prime
+ p \vec{\lambda}^{\prime T} (M_1^T)^{-1}\vec{k}
\right),
\nonumber \\
\vec{\rho}(p,\vec{\lambda}) & = & -p(I+\phi)\vec{k} +  (M_1^T)^{-1}\vec
{\lambda}.
\nonumber
\eea

The matrix notation we mentioned in the introduction can be now
put forward. We can consider the string state as an operator
acting on one copy of the Hilbert space corresponding to the half-string,
its explicit form can be extracted from the states (\ref{COHERENT2}).
Alternatively we can define the matrix elements taking the scalar
product with a generic string state with definite ``comma" occupation numbers.
The associated matrix is then defined as follows:
\be
[\vec{\lambda},\vec{\lambda}^\prime]
^{\{n_i^{(1)}\}^\infty_{i=1}}
_{\{n_i^{(2)}\}^\infty_{i=1}} =
(-)^{\sum^\infty_{i=1}n_i^{(2)}}
\langle\{n_i^{(1)}\} ; \{n_i^{(2)}\} \|
\vec{\lambda},\vec{\lambda}^\prime),
\label{MAT1}
\ee
where $|\{n_i^{(1)}\} ; \{n_i^{(2)}\}\rangle$ is the tensor product of two
``comma" states, namely
\be
|\{n_i^{(r)}\}^\infty_{i=1}\rangle =
\prod^\infty_{i=1}
\frac{1}{\sqrt{n_i^{(r)}!}} (b_i^{(r)\dagger}
)^{n_i^{(r)}}\vac_r.
\label{COMMAS}
\ee
The factor $(-)^{\sum^\infty_{i=1}n_i^{(2)}}$ appears
in the definition of (\ref{MAT1}) to conform with the
standard convention in string field theory that the parametrisation of the
second half of the string is reversed.

By using standard techniques, the creation and annihilation operators
can be dealt with to work out the explicit form of the matrix
elements in (\ref{MAT1}). The final answer is:
\bea
[\vec{\lambda},\vec{\lambda}^\prime]
^{\{n_i^{(1)}\}^\infty_{i=1}}
_{\{n_i^{(2)}\}^\infty_{i=1}} =
e^{ip{\bf x(\spin)}}
{\bf C}(p,\vec{\lambda}^\prime)
\prod^\infty_{i=1} \frac{1}{\sqrt{n_i^{(1)}! n_i^{(2)}!}}
\left(
-\frac{1}{\sqrt{2}} D^-_i
\right)
^{n_i^{(1)}}
\nonumber \\
\left(
\frac{1}{\sqrt{2}} D^+_i
\right)
^{n_i^{(2)}}
e^{-\frac{1}{2}\vec{z}^T\phi \vec{z}}
\mid_{\vec{z}=0}.
\label{MAT2}
\eea
The only new quantities are $D^\pm$ containing derivatives with respect to the
auxiliary parameter $\vec{z}$.  They are defined by:
\bea
D^\pm_i = \frac{\partial}{\partial z_i} - pk_j(I + \phi)_{ji} +
\lambda^\prime_j  (M_1^T)^{-1}_{j,i} \pm \lambda_i,
\nonumber
\eea

This expression gives the matrix form of the coherent state basis of the
string representation of the physical string states.  It can be used in
equations like (\ref{INTERACTION}) to express the star product
as a product of matrices and integration as a trace.
This will be described in detail in the next section.

Before going on, two comments are in order. Firstly, notice that
the matrix elements defined in equations (\ref{MAT1},\ref{MAT2})
can be viewed as the ones corresponding to the change of basis between the
representations given by the string states (\ref{COHERENTFULL})
and the ``comma" states (\ref{COMMAS}).
If the Hilbert space were finite dimensional, the transformation would
be automatically complete since, by construction, it relates two
orthogonal bases of it. In our case,
the string Hilbert space being of infinite dimension, one should actually
prove this property explicitly by working out Parseval's
identity both ways. This proof was
carried out in \cite{JOSE} showing the equivalence between the two
representations.

Secondly, in order to establish the validity of the ``comma" representation to
describe String Field Theory we should be able to generate the physical state
spectrum and the scattering amplitudes among the string states.
To that end we have constructed in \cite{JOSE} the Virasoro algebra in
the half-string representation,
taking care of the possible ambiguities coming from normal ordering
in both representations. It was shown that gauge and on-shell
conditions are kept in the ``comma" representation.

\section{THE N-STRING INTERACTION VERTEX}

This section is devoted to the calculation of the generic N-String
vertex. Once the matrix representation of the string states has been
put forward, we can interpret the $\star$ and integral (\ref{STAR},
\ref{WINT}) in terms of
these matrices. In fact, since the $\star$-product involves
the identification between left and right parts of contiguous
strings, we immediately see that this translates into our picture
as the product of matrices representing two contiguous strings.
On the other hand the integral, which in some sense closes the cycle, is just
the trace of the product of these matrices.

Therefore, defining the string states
through the coherent states $|\vec{\lambda}^{(k)},
\vec{\lambda}^{(k)\prime})$ given in (\ref{COHERENT1}), the N-string
interaction vertex, according to (\ref{INTERACTION}) and the
comments in the paragraph above, reads:
\bea
{\bf \cal V}_N= \int d{\bf x} \left( \spin \right)
\; e^{ \sum_{i=1}^{N} \, p^{(i)} \, {\bf x} \left( \spin \right) }
\;
\Tr
\left(
[\vec{\lambda}^{(1)},\vec{\lambda}^{(1)\prime}]
\ldots
[\vec{\lambda}^{(N)},\vec{\lambda}^{(N)\prime}]
\right),
\nonumber
\eea
where the matrices $[\vec{\lambda}^{(k)},\vec{\lambda}^{(k)\prime}]$ were
defined in (\ref{MAT2}).  The midpoint coordinate has been explicitly
separated. Integration over this coordinate can be performed in a
straightforward way, giving a $\delta$-function of conservation of momentum.
This proves that the midpoint plays the role of
the translational mode of the string as one would guess from the definition
of the ``comma" coordinates.
Thus, the part in the vertex involving the trace of the product of
matrices contains all the relevant information. It is given explicitly by:
\bea
\lefteqn{
\Tr
\left(
[\vec{\lambda}^{(1)},\vec{\lambda}^{(1)\prime}]
\ldots
[\vec{\lambda}^{(N)},\vec{\lambda}^{(N)\prime}]
\right)=
}
\nonumber \\
& &
=
\prod_{i=1}^N
{\bf C} (p^{(i)}, \vec{\lambda}^{(i)\prime})
\sum_{\{n_i^{(r)}\}}
\; \; \prod^\infty_{i=1}
\frac{1}{n_i^{(1)}! \ldots n_i^{(N)}!}
\nonumber \\
& &
\left(
-\frac{1}{2} D^{+(1)}_i D^{-(N)}_i
\right)
^{n_i^{(1)}}
\left(
-\frac{1}{2} D^{+(2)}_i D^{-(1)}_i
\right)
^{n_i^{(2)}}
\ldots
\left(
-\frac{1}{2} D^{+(N)}_i D^{-(N-1)}_i
\right)
^{n_i^{(3)}}
\nonumber \\
& &
\exp
\left(
-\frac{1}{2}
\vec{{\bf z}}
{\bf \Phi}
\vec{{\bf z}}
\right)
\mid_{\vec{{\bf z}}=0} ,
\nonumber
\eea
where the upper index in $D^{(r)}_i$ refers to the $r$-th string state,
$\vec{{\bf z}} = (\vec{\bf z}^{(1)}, \ldots,
\vec{\bf z}^{(N)})$ and ${\bf \Phi} = \phi\,{\bf I}_N$,
${\bf I}_N$ being the identity matrix
in the $N$ dimensional space spanned by the $N$
strings.

The sum over $n_i^{(r)}$ can be readily performed giving a more manageable
expression:
\bea
{\bf \cal V}_N & = &
\delta\!
\left(
\sum^N_{r=1} p^{(r)}
\right)
\prod^N_{i=1}
{\bf C}(p^{(1)},\vec{\lambda}^{(1)\prime})
\ldots
{\bf C}(p^{(N)},\vec{\lambda}^{(N)\prime})
\nonumber \\
& &
\exp
\left(
-\frac{1}{2}
\left[
\vec{D}^{-(1)} \vec{D}^{+(N)} + \vec{D}^{-(2)}
\vec{D}^{+(1)} + \ldots +  \vec{D}^{-(N)} \vec{D}^{+(N-1)}
\right]
\right)
\nonumber \\
& &
\exp
\left(
-\frac{1}{2}\vec{{\bf z}}{\bf \Phi}\vec{{\bf z}}
\right)
\mid_{\vec{{\bf z}}=0}.
\nonumber
\eea

The derivatives in the auxiliary variable
${\bf \vec{z}}$ can be carried out by using
standard techniques on quadratic forms and after a tedious, although
straightforward, calculation one ends up  with the expression for the vertex.
Up to a global normalization factor we have:
\bea
{\bf \cal V}_N =
\delta\!
\left(
\sum_{r=1}^N p^{(r)}
\right)
\exp
\left(
b\sum^N_{r=1} p^{(r)^2}
\right)
\exp
\left(
\vec{{\bf \Lambda}}^T {\bf B}_1 \vec{{\bf \Lambda}} +
\vec{{\bf \Lambda}}^{\prime T} {\bf B}^\prime_1 \vec{{\bf \Lambda}}^\prime
\right)
\nonumber \\
\exp
\left(
\vec{{\bf \Lambda}}^T {\bf B}_2 \vec{{\bf p}} +
\vec{{\bf \Lambda}}^{T \prime} {\bf B}_2^\prime \vec{{\bf p}} +
\vec{{\bf \Lambda}}^T {\bf B} \vec{{\bf \Lambda}}^\prime
\right).
\label{VERTEX}
\eea

In (\ref{VERTEX}), bold face characters refer to vectors and matrices in
the $N$-space spanned by the $N$ strings, namely
$\vec{{\bf p}} = (p^{(1)} \vec{k}, \ldots, p^{(N)} \vec{k})$,
$\vec{{\bf \Lambda}} =(\vec{{\bf \lambda}}^{(1)}, \ldots,
\vec{{\bf\lambda}}^{(N)})$, and
$\vec{{\bf\Lambda}}^{\prime}=(\vec{{\bf\lambda}}^{(1)\prime}, \ldots,
\vec{{\bf\lambda}}^{(N)\prime})$, on the other hand,
${\bf B}$,
${\bf B_i}$, and
${\bf B_i^\prime}$
are $N \times N$ dimensional matrices
whose elements are again infinite dimensional matrices.

These quantities have the following explicit
expressions in terms of the matrices of change of representation
$M_1$ and $M_2$:
\bea
b & = & -\frac{1}{2}{\bf \vec{k}^T}
\left( {\bf M_1^T} + {\bf M_2^T} \right)
\left( 1+{\bf S_+} \right)
\left( {\bf M_1^T} - {\bf S_+} {\bf M_2^T} \right)^{-1}
{\bf \vec{k}},
\nonumber \\
{\bf B}_1  &=&
\frac{1}{2}
\left[ {\bf S_+} -
{\bf S_-^T }
\left( {\bf M_1^T} - {\bf S_+} {\bf M_2^T} \right)^{-1}
{\bf M_2} {\bf S_-}
\right],
\nonumber \\
{\bf B}^\prime_1 & = &
\frac{1}{2}
\left( {\bf M_1^T} - {\bf S_+} {\bf M_2^T} \right)^{-1}
\left( {\bf M_2^T} - {\bf S_+} {\bf M_1^T} \right),
\nonumber \\
{\bf {\vec B}}_2 & = &
{\bf S_-} \left( {\bf M_1^T} - {\bf S_+} {\bf M_2^T} \right)^{-1}
\left( {\bf M_2} + {\bf M_1} \right){\bf \vec{k}},
\nonumber \\
{\bf {\vec B}}^\prime_2 & = &
- \left( {\bf M_1^T} - {\bf S_+} {\bf M_2^T} \right)^{-1}
\left( 1 + {\bf S_+ } \right) {\bf \vec{k}},
\nonumber \\
{\bf B} & = &
- {\bf S_-}
\left( {\bf M_1^T} - {\bf S_+} {\bf M_2^T} \right)^{-1},
\label{MATRICES2}
\eea
(the only quantities not yet defined
are the matrices $({\bf S_\pm})_{ij} = \frac{1}{2}
\left(\delta_{i+1,j} \pm \delta_{i-1,j}\right) $, where the lower indices are
defined
$mod \; N$).

We first diagonalize the matrix ${\bf S_+}$. It is
worthwhile going into some details. The characteristic equation
for ${\bf S_+}$ (det$\left[ {\bf S_+} - \lambda {\bf I_N} \right] = 0$)
can be written as:
$$
det( {\bf S_{+} - \lambda  {\bf I_{N}}})
= -\lambda  {\tt det} {\bf M_{N-1}} - 2  {\tt det} {\bf M_{N-2}}
-2 (-)^N=0,
$$
where $ {\bf M_N} $ is the $N$-dimensional matrix with
elements $ ({\bf M_N})_{i,j} = \frac{1}{2} (\delta_{i,j+1} +
\delta_{i,j-1}) $.
The determinant $det{\bf M_N}$, and in turn the above
equation, can be calculated
in a recursive form in terms of the dimension $N$. We can write this
in matrix form:
\bea
\left(
\begin{array}{c}
{\bf M_N} \\ {\bf M_{N-1}}
\end{array}
\right) =
\left(
\begin{array}{cc}
-2 \lambda & -1 \\
1 & 0
\end{array}
\right)
\left(
\begin{array}{c}
{\bf M_{N-1}} \\ {\bf M_{N-2}}
\end{array}
\right),
\eea
with the subsidiary conditions det${\bf M_{1}}=-\lambda$
and  det${\bf M_{0}}=1$. After diagonalization
one gets the solution for  det${\bf M_{N}}$:
$$
det{\bf M_{N}}= \frac{\mu^{N+1}_+ - \mu^{N-1}_-}{\mu_+ - \mu_-},
$$
where $\mu_{\pm}= -\lambda \pm \left[ \lambda^2 -1 \right]^{1/2}$.
Hence, the solutions of the characteristic equation fulfills:
$$
\left[ -\lambda \pm \left( \lambda^2 -1 \right)^{1/2}
\right]^N = (-)^N.
$$
Solving this equation we get the values of $\lambda = \cos \frac{2k\pi}{N}$
with $k=1 \ldots N$ which are the eigenvalues of the matrix
${\bf S_+}$.

Now, the matrix that performs the diagonalization of ${\bf S_+}$
(we use the notation ${\bf S_+} = {\bf R^T} {\bf D} {\bf R}$,
with ${\bf D}_{i,j}= \delta_{i,j} \cos \frac{2k\pi}{N}$)
is given by:
\bea
{\bf R}_{i,j} & = & \sqrt{\frac{2}{N}} \cos \frac{\pi i}{N}(2 j - 3),
\; \; \; {\tt for} \; \; \;
1 \leq k \leq  \left[ \frac{N-1}{2} \right] ,
\nonumber \\
{\bf R}_{N/2,j} & = & \sqrt{\frac{1}{N}} (-)^{j+1},
\nonumber \\
{\bf R}_{i,j} & = & \sqrt{\frac{2}{N}} \sin \frac{\pi i}{N}(2 j - 3),
\; \; \; {\tt for} \; \; \;
\left[ \frac{N+1}{2} \right] < k < N,
\nonumber \\
{\bf R}_{N,j} & = & \sqrt{\frac{1}{N}}.
\label{ROTATION}
\eea

Substituting in (\ref{MATRICES2}) and using the symmetry properties
of the trigonometric functions one ends up with the following expressions
for the terms appearing in the exponent of the vertex:
\bea
b & = & -\frac{4}{N} \sum^{\left[ \frac{N-1}{2} \right]}_{k=1}
\cos^2 \frac{k \pi}{N} \cos \frac{2k \pi (i-j)}{N}
\nonumber \\
& &
\vec{k}^T \left( M_1^T + M_2^T \right)
\left( M_1^T - \cos \frac{2 k \pi}{N} M_2^T \right)^{-1}
\vec{k},
\nonumber \\
{\bf B}^{i,j}_1  &=&
\frac{1}{2} \left( \delta_{i,j+1} +  \delta_{i,j-1} \right) I-
\nonumber \\
&- &
\frac{2}{N} \sum^{\left[ \frac{N-1}{2} \right]}_{k=1}
\sin^2 \frac{2 k \pi}{N} \cos \frac{2k \pi(i-j)}{N}
M_2^T \left( M_1^T - \cos \frac{2 k \pi}{N} M_2^T \right)^{-1},
\nonumber \\
{\bf B}_1^{\prime i,j} & = &
\frac{1}{N} \left[ (-)^{i+j} - 1 \right] I +
\nonumber \\
& + &
\frac{2}{N} \sum^{\left[ \frac{N-1}{2} \right]}_{k=1}
\cos \frac{2k \pi(i-j)}{N}
\left( M_1^T - \cos \frac{2 k \pi}{N} M_2^T \right)^{-1}
\left( M_2^T - \cos \frac{2 k \pi}{N} M_1^T \right),
\nonumber \\
{\bf {\vec B}}_2^{i,j} & = &
\frac{2}{N} \sum^{\left[ \frac{N-1}{2} \right]}_{k=1}
\sin \frac{2 k \pi}{N} \sin \frac{2k \pi(i-j)}{N}
\left( M_1^T - \cos \frac{2 k \pi}{N} M_2^T \right)^{-1}
\left( M_2 + M_1 \right)\vec{k},
\nonumber \\
{\bf {\vec B}}^{\prime i,j}_2 & = &
\frac{2}{N} \left( M_1 + M_2 \right)\vec{k} +
\nonumber \\
&+ &
\frac{4}{N} \sum^{\left[ \frac{N-1}{2} \right]}_{k=1}
\cos^2 \frac{k \pi}{N} \cos \frac{2k \pi(i-j)}{N}
\left( M_1^T - \cos \frac{2 k \pi}{N} M_2^T \right)^{-1}
\vec{k},
\nonumber \\
{\bf B}^{i,j} & = &
\frac{2}{N} \sum^{\left[ \frac{N-1}{2} \right]}_{k=1}
\sin \frac{2 k \pi}{N} \sin \frac{2k \pi(i-j)}{N}
\left( M_1^T - \cos \frac{2k \pi}{N} M_2^T \right)^{-1}.
\label{MATRICES3}
\eea

This gives the final form of the vertex obtained in this approach.
The important
thing to notice at this point is that this result is given in terms of the
change of representation matrices $M_1$ and $M_2$ (\ref{MATRICES}), that appear
in the combination
$\left( M_1^T - \cos \frac{2k\pi}{N} M_2^T \right)^{-1}$
with $k= 1 \ldots N$. Once the inverse of this matrix is obtained, we can
directly identify the elements of the exponent of equation (\ref{VERTEX}) with
the Fourier components of the Neumann function corresponding to the $N$-string
contact interaction. We postpone the calculation of this matrix and the
identification with the Neumann functions until the next section.

Before concluding this section let us make some comments on the symmetries of
the coefficients appearing in (\ref{MATRICES3}).
First notice that regarding to
the diagonal terms ($i=j$), they are independent of the position. This
corresponds to the fact that they describe the self interaction of the
strings and, in the general form worked here, they are equivalent.
The second property to notice is that, if $A$ is any of the matrices
in (\ref{MATRICES3}), they fulfill
$A^{i,j}=A^{i+l,j+l}$ as it corresponds to the fact that they describe
the interaction of the string in position $i$ ($i+l$) with
the one at position $j$ ($j+l$). Finally the diagonal terms
in the matrices ${\bf B}^{i,j}$
vanish, this corresponds to the lack of connection
between $odd$ and $even$ modes in the same string.

\section{NEUMANN COEFFICIENTS FOR THE N-STRING VERTEX}

In this section we finish the calculation of the vertex and proceed
to the identification of the coefficients appearing in the
exponent of the vertex in equations (\ref{VERTEX}) and (\ref{MATRICES3}),
with the ones
obtained via the conventional approach based on the path integral formulation
of string amplitudes. We want to stress the fact that in this approach
the Neumann coefficients are obtained in a compact form once the
way to invert the matrix
$\left( M_1^T - \cos \frac{2k\pi}{N} M_2^T \right)^{-1}$
is known.

Before going on let us outline briefly the calculation of the
vertex based on the path integral formalism of the string sigma model.

The $N$-string amplitude for a closed region is given by the path
integral:
\bea
\int D X(z)
exp \left\{
- \frac{1}{4 \pi} \int d^2 z \; \; \left( \partial X(z) \right)^2
+ \sum_{r=1}^N \int^{\pi}_0 dz_r \; \; X \left( z_r \right)
p^r\left( z_r \right) \right\},
\nonumber
\eea
where a general state is represented by the momentum distribution
$p^r(z_r)$. The vertex is given by:
\bea
{\cal V}_N =
exp \left\{
-\frac{1}{2} \sum_{r,s=1}^N
\int_0^\pi dz_r  \int_0^\pi dz_s \; \;
p^r(z_r) N(z_r,z_s)  p^r(z_s)
\right\},
\nonumber
\eea
where the Neumann function is defined as the solution of the differential
equation:
\bea
\left( \partial^2_\tau + \partial^2_\sigma \right)
N(z,z^\prime)
&=&
2 \pi \delta^2 (z-z^\prime),
\nonumber \\
\partial_n N(z,z^\prime)
& = &
f(z),
\nonumber
\eea
where $\vec{n}$ is a vector normal to the boundary.
The former equation can be cast in terms of the Fourier components of
the Neumann function as:
\bea
{\cal V}_N =
exp \left\{
\frac{1}{2} \sum_{r,s=1}^N \sum_{n,m=1}^\infty
p^r_n N_{n,m} p^r_m.
\right\}.
\label{VERTEXNEUMANN}
\eea

Now we can compare the vertices of equations (\ref{VERTEX}) and
(\ref{VERTEXNEUMANN}) and identify the quantities in (\ref{MATRICES3})
with the Fourier components of the Neumann functions.

To get an explicit expression, we proceed to invert the matrix
$$
\left( M_1^T - \cos \frac{2k\pi}{N} M_2^T \right)^{-1}.
$$
Consider the generic combination:
$$
\alpha M_2^T - \beta M_1^T,
$$
we propose  the ansatz:
\bea
\alpha^\prime
\frac{ v^{(1/p)}_{2m} u^{(1/p)}_{2n-1}+
v^{(1/p)}_{2n-1} u^{(1/p)}_{2m}}{2m-2n+1} +
\beta^{\prime}
\frac{u^{(1/p)}_{2n-1} v^{(1/p)}_{2m}
- v^{(1/p)}_{2n-1} u^{(1/p)}_{2m}}{2m+2n-1}
\nonumber
\eea
for the inverse.
The coefficients $u_n^{(1/p)}$ and $v_n^{(1/p)}$ are the
coefficients of the Taylor expansion of the rational functions:
\bea
\left(
\frac{1+x}{1-x}
\right)^{1/p}, \; \; \; \;
\left(
\frac{1+x}{1-x}
\right)^{1-1/p}
\nonumber
\eea
respectively.
On the other hand $\alpha^\prime$ and $\beta^\prime$ as well as $p$
are free parameters to be determined.

Imposing the condition that our ansatz is the required inverse
we
end up with the following equations that restrict the values of the
free parameters $\alpha^\prime$, $\beta^\prime$ and $1/p$:
\bea
\alpha \alpha\prime - \beta \beta^\prime \cos \frac{\pi}{p} & = & 0,
\nonumber \\
\alpha \beta - \beta \alpha^\prime \cos \frac{\pi}{p} & = & 0,
\nonumber \\
\frac{(-)^{n+m}}{2 \sin \frac{\pi}{p}} (2m)^{-1/2}  (2n)^{-1/2}
& = &  \beta \alpha^\prime .
\nonumber
\eea
{}From these equations we fix the free parameters.
The relations read:
\bea
\alpha^\prime & = &
- \frac{1}{ 4 } \frac{1}{\sin \frac{\pi}{p} \beta},
\nonumber \\
\beta^\prime & = &
\frac{1}{4}\frac{\cos \frac{\pi}{p}}{\sin \frac{\pi}{p}\beta},
\nonumber \\
\cos^2 \frac{\pi}{p} & = &
\frac{\alpha^2}{\beta^2}.
\eea
In the particular case of interest to us, the values of the
free parameters are:
$$
\alpha=\cos \frac{2 k \pi}{N},
\; \; \;
\beta=1
\; \; \; {\tt and} \; \; \;
cos^2\frac {\pi}{p} =\cos^2 \frac{2 k \pi}{N}.
$$
{}From them we obtain:
\bea
\alpha^\prime  =  \beta^\prime & = & -\frac{1}{4 \sin \frac{2 k \pi}{N}}
\; \; \; {\tt and}\; \; \;
p = \frac{N}{2k}
\; \; \; {\tt for}\; \; \;
k=1...\left[\frac{N-1}{2}\right],
\nonumber \\
{\tt or} \; \; \;
p & = & \frac{N}{2(N-k)}
\; \; \; {\tt if} \; \; \;
k=\left[\frac{N+1}{2}\right]...(N-1)
\nonumber
\eea
(the cases $k=N,\frac{N}{2}$ are trivially solved).
These results complete the form of the inverse matrix.
For instance if $k<\left[\frac{N-1}{2}\right]$ the result reads:
\bea
& &\left( M_1^T - \cos \frac{2k\pi}{N} M_2^T \right)^{-1} \mid_{m,n}
=
\nonumber \\
& = & \frac{(-)^{n+m} (2m)^{1/2} (2n-1)^{1/2}}{2 \sin \frac{2 k \pi}{N}}
\left[
\frac{ v^{(2k/N)}_{2m} u^{(2k/N)}_{2n-1}+
v^{(2k/N)}_{2n-1} u^{(2k/N)}_{2m}}{2m-2n+1} +
\right.
\nonumber \\
& & +
\left.
\frac{u^{(2k/N)}_{2n-1} v^{(2k/N)}_{2m}
- v^{(2k/N)}_{2n-1} u^{(2k/N)}_{2m}}{2m+2n-1}
\right],
\nonumber
\eea

Once the inverse matrix is obtained, it is a straightforward matter
to calculate the matrices appearing in (\ref{MATRICES3})
and the further identification with the Neumann functions.
To illustrate the sort of calculation involved,
let us consider the case of the
quadratic momentum term. We want to compute the quantity:
\bea
\vec{k}^T \left( M_1^T + M_2^T \right)
\left( M_1^T - \cos \frac{2k\pi}{N} M_2^T \right)^{-1}
\vec{k},
\nonumber
\eea
which can be written as:
\bea
\frac{8}{\pi^3} \frac{1}{\sin \frac{2k\pi}{N}}
\frac{1}{(2n-1)^2}
\left( \frac{1}{2m+2n-1} + \frac{1}{2m-2n+1} \right)
v^{(2k/N)}_{2m} \Sigma^u_0.
\nonumber
\eea
Then, using the properties of the sums given in the appendix --
(\ref{SUMSODD}),  (\ref{SUMSNEGATIVE}) and
(\ref{SIGMATILDE0}) in particular--
we end up with the result:
$$
\frac{2}{\pi \sin \frac{2k\pi}{N}} \tilde{\Sigma}^v_0,
$$
which leads to the value of the Neumann coefficient $N_{0,0}$.
Proceeding in a similar fashion we can obtain the remaining
Neumann coefficients described in
(\ref{MATRICES3}) and  (\ref{VERTEXNEUMANN}). The final result is:
\bea
& &
N_{0,0}^{i,j} =
\frac{1}{N} \sum^{\left[ \frac{N-1}{2} \right]}_{k=1}
\cos \frac{2k \pi(i-j)}{N}
\left[
\psi \left( 1- \frac{k}{N} \right) +
\psi \left( \frac{k}{N} \right) -
2 \psi(1) + 4 \ln 2
\right],
\nonumber \\
& & (-)^n N_{n,m}^{i,j} =
-\delta_{n,m} \frac{\delta_{i,j-1}+ \delta_{i,j+1}}{2}+
\frac{2}{N} \sum^{\left[ \frac{N-1}{2} \right]}_{k=1}
\cos \frac{2k \pi(i-j)}{N}
\frac{(-)^{n+m}}{2} n^{1/2} m^{1/2},
\nonumber \\
& &
\left[
\frac{u^{(\frac{2k}{N})}_{n} v^{(\frac{2k}{N})}_{m} +
v^{(\frac{2k}{N})}_{n}  u^{(\frac{2k}{N})}_{m}}
{n + m} +
\frac{u^{(\frac{2k}{N})}_{n} v^{(\frac{2k}{N})}_{m} -
u^{(\frac{2k}{N})}_{m}  v^{(\frac{2k}{N})}_{n}}
{n - m}
\right],
\; \; \;  {\tt for} \; \; \; n+m:even,
\nonumber \\
& &
N_{n,m}^{i,j} =
-\frac{2}{N} \sum^{\left[ \frac{N-1}{2} \right]}_{k=1}
\sin \frac{2k \pi(i-j)}{N}
\frac{(-)^{n+m}}{2} n^{1/2} m^{1/2}
\nonumber \\
& &
\left[
\frac{u^{(\frac{2k}{N})}_{n} v^{(\frac{2k}{N})}_{m} +
v^{(\frac{2k}{N})}_{n}  u^{(\frac{2k}{N})}_{m}}
{n - m} -
\frac{u^{(\frac{2k}{N})}_{n} v^{(\frac{2k}{N})}_{m} -
u^{(\frac{2k}{N})}_{m}  v^{(\frac{2k}{N}}_{n})}
{n + m}
\right],
\; \; \; {\tt for} \; \; \; n+m:odd,
\nonumber \\
& &
N_{0,2n}^{i,j} =
\frac{2}{N} \sum^{\left[ \frac{N-1}{2} \right]}_{k=1}
\cos \frac{2k \pi(i-j)}{N}
\frac{(-)^n}{(2n)^{1/2}}
v^{(\frac{2k}{N})}_{2n},
\nonumber \\
& &
N_{0,2n-1}^{i,j} =
\frac{2}{N} \sum^{\left[ \frac{N-1}{2} \right]}_{k=1}
\sin \frac{2k \pi(i-j)}{N}
\frac{(-)^n}{(2n-1)^{1/2}}
v^{(\frac{2k}{N})}_{2n-1}.
\label{FINALRESULT}
\eea
Which is only valid for $N\geq 2$ (when $N=2$, there are no
sums).
These equations do not include the case $N=1$, although it can be treated
trivially in the half-string formulation, see for example \cite{GHOSTS}.
This completes the calculation of the N-sting interaction
vertex (\ref{VERTEX}) and represents our main result.
The interaction between string Fock space states is readily obtained by
taking
derivatives with respect to
the parameters $\lambda$, $\lambda^\prime$,
as it was indicated in (\ref{COHERENT2}).

We insist again on the fact that the Neumann functions are generated in an
explicit way from the representation changing matrices $M_1$ and $M_2$.
Therefore, in this picture they appear as derived quantities.  This assertion
is true for every $N$-string contact interaction, thus providing a relation
among the Neumann functions associated with the $N$-string vertices.

To illustrate these results we can consider the case
$N=3$, for which the sums in (\ref{FINALRESULT}) only have the term
corresponding to $k=1$. The result one obtains from  (\ref{FINALRESULT})
agrees with previous calculations of the cubic interaction vertex
performed earlier \cite{GROSS,MCARTHY}.

\section{CONCLUSIONS}

In this work we have calculated the vertex corresponding to the contact
interaction of $N$ strings. We have used the ``comma" representation of
String Field theory in which the prominent role that
the joining of the half strings plays is apparent. This feature also
appeared in the pioneering work of
Witten \cite{WITTEN}.

This approach has the advantage of furnishing a
compact treatment of the vertex giving
a result of general validity, independent of the number of strings.
The final answer is always given in terms
of a matrix which involves particular combinations between the matrices
that change from the string representation (in which string physical states
take on a simple form) to the ``comma" representation in which
interaction takes a trivial form.

The relevant matrix has been calculated and in general is given in terms
of the Taylor coefficients of particular rational functions. Those
coefficients, their sum rules and most of their properties which are
relevant to our
work, have been worked out in the appendix. With this result one
can readily identify the Fourier coefficients of the Neumann
function for the $N$-string geometry. Finally we checked our results
against
the simple
case of the cubic interaction in agreement with the known
results.

The task to face know is two folded. On the one hand, work is under
way to implement this result to the case of closed strings. The required
modifications are trivial and will be reported in the near future.
On the other hand, the extension to the description of the restricted
polyhedra describing the terms in non-polynomial action of closed strings
will require the study of these vertices under reparametrizations.
Work in this direction is under way.
In particular it is no difficult to prove \cite{FUTURO} that the modular
parameters appearing in the CSFT, namely the length of the
string overlaps, are related in a straightforward way
to the reparametrization parameters of \cite{CHANII}.

{\bf Acknowledgments}

The work of J.B. has been supported by CICYT (Spain) under
Grant AEN-90-0040. F.A. wants to thank
UNAM (M\'exico) for their support. The work of A.A.
has been supported by SSR (Libya). The authors thank H. M. Chan
for his valuable comments.

\newpage

\appendice

In this appendix we give the properties of the coefficients
of the Taylor expansion of the functions:
\bea
\left(
\frac{1+x}{1-x}
\right)^{1/p} & = &
\sum_{n=1}^{\infty} \;
u^{(1/p)}_n x^n ,
\nonumber \\
\left(
\frac{1+x}{1-x}
\right)^{1-1/p}
& = &
\sum_{n=1}^{\infty} \;
v^{(1/p)}_n x^n,
\nonumber
\eea
we need them for the construction of the Fourier coefficients of the
Neumann functions carried out in section 5.

Most of the results derived here are a generalization of the
calculations performed in ref. \cite{GROSS,MCARTHY} where the case
$p=3$ was analyzed in detail.

With the above definition of the coefficients $u^{(1/p)}_n$ and
$v^{(1/p)}_n$, we can express them in an integral form:
\bea
u^{(1/p)}_n & = &
\frac{1}{2 \pi i} \oint_0
\frac{dx}{x^{n+1}}
\left(
\frac{1+x}{1-x}
\right)^{1/p} .
\nonumber
\eea
The $v^{(1/p)}_n$'s are expressed in an analogous fashion
with the substitution of
$1/p$ by $1-1/p$. This expression is useful to find the
recursion relation obeyed by the coefficients. Integration
by parts of the derivative of the integrated function gives:
\bea
\frac{2}{p} u^{(1/p)}_n & = &
(n+1) u^{(1/p)}_{n+1} - (n-1) u^{(1/p)}_{n-1} ,
\nonumber \\
2 \left( 1- \frac{1}{p} \right)
v^{(1/p)}_n & = &
(n+1) v^{(1/p)}_{n+1} - (n-1) v^{(1/p)}_{n-1}.
\label{RECURSION1}
\eea

Also, making use of the same integral representation,
one can relate both coefficients:
\bea
\frac{2}{p} v^{(1/p)}_n & = &
(-)^n \left[ (n+1) u^{(1/p)}_{n+1} - 2 n u^{(1/p)}_n + (n-1) u^{(1/p)}_{n-1}
\right] ,
\nonumber \\
2 \left( 1 - \frac{1}{p} \right) u^{(1/p)}_n & = &
(-)^n \left[ (n+1) v^{(1/p)}_{n+1} - 2 n v^{(1/p)}_n + (n-1) v^{(1/p)}_{n-1}
\right] .
\label{UNANDVN}
\eea

It is as well possible to find a closed expression for the
coefficients although it wont be necessary for our purposes.

In the text, and in particular in section 5 one needs
the evaluation of several infinite
sums involving these coefficients.
The simplest of these sums is:
\bea
\Sigma_n^u = \sum_{n+m:odd} \frac{u^{(1/p)}_n}{n+m}.
\nonumber
\eea
It can be written in an integral form allowing its evaluation.
For instance, for $odd$ indices, we have:
\bea
\sum_{m=1} \frac{u^{(1/p)}_{2m-1}}{2n+2m-1} =
\frac{1}{\sin \frac{\pi}{p}}\frac{1}{8i}
\oint_0 \frac{dx}{x^{2n+1}}
\left[ \left( \frac{1+x}{1-x} \right)^{1/p} +
\left( \frac{1+x}{1-x} \right)^{-1/p}
\right],
\nonumber
\eea
the last integral being proportional to the
coefficient $u^{(1/p)}_{2n}$.
As a result we find:
\bea
\Sigma_n^u = \sum_{n+m:odd} \frac{u^{(1/p)}_m}{n+m}
= \frac{\pi}{2} \frac{1}{\sin \frac{\pi}{p}} u^{(1/p)}_n.
\label{SUMSODD}
\eea

{}From this relation, it is then straightforward
to deduce a recursion relation for these sums
similar to the one found previously for the coefficients $u^{(1/p)}_n$ and
 $v^{(1/p)}_n$
(\ref{RECURSION1}):
\bea
\frac{2}{p} \Sigma^u_n = (n+1) \Sigma^u_{n+1}
- (n-1) \Sigma^u_{n-1}.
\label{RECURSIONSUMS}
\eea

We can extrapolate this result to find the sums for negative
values of the index, namely:
\bea
\Sigma_{-n}^u = - \frac{\pi}{2} \cot \frac{\pi}{p} u^{(1/p)}_n.
\label{SUMSNEGATIVE}
\eea

To find this result we need a boundary condition that can
be given by the sum $\Sigma_0$. We can
proceed by direct integration:
\bea
\Sigma_0^u & = & \frac{1}{2}
\int_1^\infty \frac{dx}{x}
\left[
\left( \frac{x+1}{x-1} \right)^{1/p} -
\left( \frac{x-1}{x+1} \right)^{1/p}
\right] =
\nonumber \\
& = & \frac{1}{2}  \left[ \psi \left( \frac{1}{2} + \frac{p}{2} \right)
- \psi \left( \frac{1}{2} - \frac{p}{2}  \right) \right] =
\frac{\pi}{2} \tan \frac{\pi}{2p}.
\nonumber
\eea

The change of variable one needs to perform the integration,
namely $y=\left(\cosh (\frac{1}{2} \ln x) \right)^{-2}$,
was already suggested in \cite{GROSS}.

Note that, in the sums $\Sigma_n$ evaluated above, one is summing
over the index $m$ with parity opposed to $n$. The sums for indices
with the same parity are more involved and we will only discuss the
properties which are more relevant to our work.

First the sums involving quadratic
denominators, namely:
\bea
\tilde{\Sigma}_n^u = \sum_{n+m:odd}
\frac{u^{(1/p)}_n}{(n+m)^2}.
\nonumber
\eea

One can show, solely making use of the recursion relations
given in equation (\ref{RECURSION1}), the following recursion
relation:
\bea
(n+1) \tilde{\Sigma}^u_{n+1} =
\frac{2}{p}  \tilde{\Sigma}^u_{n} + (n-1) \tilde{\Sigma}^u_{n-1}
+ \Sigma^u_{n+1} - \Sigma^u_{n-1},
\nonumber
\eea
which can be extended to negative values of the index $n$ and
by direct evaluation of the combination:
\bea
\cos \frac{\pi}{p} \tilde{\Sigma}_1^u -
\tilde{\Sigma}_{-1}^u = \frac{\pi}{2} \sin \frac{\pi}{p}
\Sigma_0^u S^u_n,
\nonumber
\eea
that we use as boundary condition, we find the relation between
the sums for positive and negative indices:
\bea
\tilde{\Sigma}_{-n}^u -
\tilde{\Sigma}_{n}^u
\cos \frac{\pi}{p} =
\frac{\pi}{2} \sin\frac{\pi}{p}
\Sigma_0^u S^u_n.
\nonumber
\eea

The sum $S_m= \sum_{n+m:even} \frac{u^{(1/p)}_n}{n+m}$,
which involves summing with
indices of the same parity, appears here.
It can directly be shown that they satisfy the
same recursion relation as before (\ref{RECURSIONSUMS}):
\bea
\frac{2}{p} S^u_n = (n+1) S^u_{n+1} - (n-1) S^u_{n-1}.
\nonumber
\eea

Notice at this point that a recursion relation
of this kind has
two different solutions, one proportional to the coefficients
$u^{(1/p)}_n$ which is the one given in (\ref{SUMSODD}) and the other
one, corresponding to the sums $S_n$ which behaves as
$1/n$ when $n \rightarrow 0$. The general form of the latter
can be obtained by using the generating function:
$$
S(x) = \sum_{n=1}^{\infty} \; S_n \, x^n,
$$
and convert the recursion relation into a differential equation whose
solution is given by:
$$
S(x) = \left( \frac{1+x}{1-x} \right)^{1/p}
\int_0^x \; dy \;
\left( \frac{1+y}{1-y} \right)^{1/p}
\frac{u^{(1/p)}_0 y + S_1}{1-y^2}.
$$
This can be solved in terms of the original coefficients.
We will skip the details, instead we evaluate several combinations
of these sums which are relevant in section 5.

Let us calculate the combination
given by:
\bea
u^{(1/p)}_{2n-1} S^v_{2n-1} + v^{(1/p)}_{2n-1} S^u_{2n-1}.
\eea
We start from the quantity:
$$
T_{n,m} = \frac{u^{(1/p)}_n v^{(1/p)}_m + u^{(1/p)}_m v^{(1/p)}_n}{n+m},
$$
that, using the relations (\ref{RECURSION1}, \ref{UNANDVN}),
can be shown to satisfy:
\bea
(m+1) T_{m+1,n} - (m-1) T_{m-1,n} +
(n+1) T_{m,n+1} - (n-1) T_{m,n-1} = 0 ,
\nonumber
\eea
for $n+m:odd$.
Now, taking $m \rightarrow 2m-1$, $n \rightarrow 2n$ and summing over
the index $m$, one finds the recursion relation:
\bea
(2n+1)
\left[ u^{(1/p)}_{2n+1} S^v_{2n+1} + v^{(1/p)}_{2n+1} S^u_{2n+1} \right]
=
(2n-1)
\left[ u^{(1/p)}_{2n-1} S^v_{2n-1} + v^{(1/p)}_{2n-1} S^u_{2n-1} \right]
\nonumber
\eea
which has a solution:
\bea
u^{(1/p)}_{2n-1} S^v_{2n-1} + v^{(1/p)}_{2n-1} S^u_{2n-1} = \frac{S_1}{2n-1}.
\nonumber
\eea
To determine the value of $S_1$ we proceed directly writing it in its integral
form to find $S_1=2$, a result that
is independent of the value of $\frac{1}{p}$.

Following the same steps, one can easily evaluate the
combinations:
\bea
 & &
u^{(1/p)}_{2n} S^v_{2n} + v^{(1/p)}_{2n} S^u_{2n} = \frac{2}{2n},
\nonumber \\
& &
u^{(1/p)}_{2n-1} S^v_{2n-1} + v^{(1/p)}_{2n-1} S^u_{ 2n-1} -
\left[
u^{(1/p)}_{2n-1} S^v_{-(2n-1)} - v^{(1/p)}_{2n-1} S^u_{-(2n-1)}
\right]
= -\frac{1}{2n-1},
\nonumber
\eea
which also appear in the calculation of the interaction vertex.

To take care of the indetermination appearing in the
functions $N_{n,m}^{i,j}$ in the limit when $n \rightarrow m$.
We can show that:
\bea
& & \lim_{n \rightarrow m}
\frac{u^{(1/p)}_{n} v^{(1/p)}_{m}
- v^{(1/p)}_{n} u^{(1/p)}_{m}}{2(n-m)} =
\nonumber \\
& = &
\frac{2}{\pi} \sin \frac{\pi}{p}
\left[
u^{(1/p)}_{m} \tilde{\Sigma}_m^v -
v^{(1/p)}_{m} \tilde{\Sigma}_m^u
\right].
\nonumber
\eea
This result is obtained just by writing the left hand side in an
integral form.

Finally, in the evaluation of the Neumann coefficient $N_{00}$ one needs
the sum: $\tilde{\Sigma}_0^u =
\sum_{n=1}^{\infty} \frac{u^{(1/p)}_{2n-1}}{(2n-1)^2}$, which can be
performed using its integral representation:
\bea
\tilde{\Sigma}_0^u & = &
- \lim_{k \rightarrow 0}
 \frac{1}{2} \frac{d}{dk}  \int_0^1 \frac{dx}{x}
x^k
\left[
\left( \frac{1+x}{1-x} \right)^{1/p} -
\left( \frac{1-x}{1+x} \right)^{1/p}
\right] =
\nonumber \\
& = & - \frac{1}{4}
\frac{\cos\frac{\pi}{2p}}{\sin \frac{\pi}{2p}}
\left[
\psi \left( \frac{1}{2} + \frac{1}{2p} \right) +
\psi \left( \frac{1}{2} - \frac{1}{2p} \right) -
2 \psi(1) + 4 \ln 2 \right]
\label{SIGMATILDE0}
\eea

To end this appendix notice that the discussion we have undertaken
for the coefficients $u^{(1/p)}_n$ can be translated into $v^{(1/p)}_n$ with
the only change of $1/p \rightleftharpoons 1-1/p$.


\newpage
\begin{thebibliography}{25}

\bibitem{WITTEN}
E. Witten, Nucl. Phys. {\bf B268} (1986) 253.
\bibitem{ZWIEBACH}
M . Kaku, Nucl. Phys. {\bf B267} (1986) 125. \\
B. Zwiebach, Ann. of Phys {\bf 186} (1988) 111.\\
H. Sonoda and B. Zwiebach,  {\bf B331} (1989) 592.\\
M. Saadi and B. Zwiebach, Ann. of Phys. {\bf B192} (1989) 213.\\
M. Kaku, Phys. Rev. {\bf D41} (1990) 3734. \\
L. Hua and M. Kaku, Phys. Lett. {\bf 250B} (1990) 56.
\bibitem{CHANII}
L. Nellen. Ph. D. Thesis. Department of Theoretical Physics,
Oxford University, 1991. \\
Chan Hong-Mo, Tsou Sheung Tsun, Lukas Nellen and J. Bordes,
Phys. Rev. {\bf D44} (1991) 1786.
\bibitem{CLOSED}
F. Ant\'on, A. Abdurrahman and J. Bordes, OUTP 93-10-P, submitted to
Phys. Rev. D. hep-th/9305166.
\bibitem{JOSE}
J. Bordes, Chan Hong-Mo, Tsou Sheung Tsun and Lukas Nellen,
Nucl. Phys. {\bf B351} (1991) 441.
\bibitem{GHOSTS}
A. Abdurrahman, F. Ant\'on and J. Bordes, {\it Half-String
Oscillator Approach to String Field Theory (Ghost Sector I)}
(To appear in Nucl. Phys. B).
\bibitem{GREENBOOK}
M. B. Green, J. H. Schwarz and E. Witten, {\it Superstring Theory},
Cambridge University Press, Cambridge 1987.
\bibitem{GROSS}
D. Gross and A. Jevicki, Nucl. Phys. {\bf B283} (1986) 57.
\bibitem{MCARTHY}
A. Schwimmer, E. Cremmer and C. Thorn, Phys. Lett. {\bf 179B} (1986) 57. \\
S. Samuel, Phys. Lett.{\bf 181B} (1986) 249. \\
A. Estaugh and J.G. McCarthy, Nucl Phys. {\bf B294} (1987) 845. \\
J. Bordes and F. Lizzi, Phys. Rev. Lett. {\bf 61} (1988) 278. \\
\bibitem{FUTURO}
F. Ant\'on, A. Abdurrahman and J. Bordes (work in progress).

\end{thebibliography}
\end{document}